\documentclass{nic-series}

\usepackage{braket}
\usepackage{pgfplots}
\usepackage{bm}
\usepackage{sty_jattana_tikzit}
\usepackage{qcircuit}

\begin{document} 

\title{Hybrid Quantum Classical Simulations}

\author{Dennis Willsch \inst{1} \and 
        Manpreet Jattana \inst{1,2} \and
        Madita Willsch \inst{1,3} \and
        Sebastian Schulz \inst{1,2} \and
        Fengping Jin \inst{1} \and
        Hans De Raedt \inst{1,4} \and
        Kristel Michielsen \inst{1,2,3}}

\authortoc{D. Willsch, M. Willsch, M. Jattana, S. Schulz, F. Jin, K. Michielsen}

\institute{Institute for Advanced Simulation, J\"ulich Supercomputing Centre, 
  Forschungszentrum J\"ulich, 
  52425 J\"ulich, Germany\\
    \email{\{d.willsch, m.jattana, m.willsch, se.schulz, f.jin, k.michielsen\}@fz-juelich.de}
  \and
  RWTH Aachen University, 52056 Aachen, Germany
  \and
  AIDAS, 52425 J\"ulich, Germany
  \and
  Zernike Institute for Advanced Materials, University of Groningen, Nijenborgh 4, NL-9747 AG Groningen, The Netherlands\\
  \email{deraedthans@gmail.com}
}

\maketitle

\begin{abstracts}
We report on two major hybrid applications of quantum computing, namely, the quantum approximate optimisation algorithm (QAOA) and the variational quantum eigensolver (VQE). Both are hybrid quantum classical algorithms as they require incremental communication between a classical central processing unit and a quantum processing unit to solve a problem. We find that the QAOA scales much better to larger problems than random guessing, but requires significant computational resources. In contrast, a coarsely discretised version of quantum annealing called approximate quantum annealing (AQA) can reach the same promising scaling behaviour using much less computational resources. For the VQE, we find reasonable results in approximating the ground state energy of the Heisenberg model when suitable choices of initial states and parameters are used. Our design and implementation of a general quasi-dynamical evolution further improves these results.
\end{abstracts}

\section{Introduction}
\label{willsch_sec_introduction}

Quantum computing\cite{NielsenChuang} is an emerging computer technology that uses quantum effects in the design of its computational model. There are two major paradigms in quantum computing, namely the gate-based quantum computer and the quantum annealer. 

A gate-based quantum computer is inspired by the circuit model for classical, digital computers. This means that every program is defined in terms of a sequence of fundamental operations, the so-called \emph{quantum gates}. Each quantum gate operates on the fundamental units of computation, the so-called \emph{quantum bits} or \emph{qubits}. The gate-based quantum computer executes these quantum gates step-by-step and thus updates its internal quantum state. At the end of the quantum gate sequence, the quantum state is ``measured'', which means that the quantum computer outputs, with a certain probability, one out of several classical bitstrings. The maximum number of bits in this bitstring is given by the number of qubits in the quantum computer. At the time of writing, gate-based quantum computers with approximately one hundred qubits have been manufactured\cite{randomrandomdevarmonk}.

A quantum annealer, on the other hand, tries to harness the natural evolution of a quantum system, steered by external magnetic fields, to solve an optimisation problem. Internally, it also operates on a set of qubits, which are measured at the end of a quantum annealing run. It thus also produces a bitstring as output. At the time of writing, quantum annealers with more than 5000 qubits have been manufactured\cite{dwave2020Advantage}.

Both types of quantum computers share the property that the same program can produce different bitstrings obtained after a run. The computational models are thus inherently probabilistic. This means that quantum computers can also be used as \emph{samplers}. A program is often executed multiple times to obtain a representative distribution of bitstrings. 

\begin{figure}
    \centering
    \includegraphics[scale=0.9]{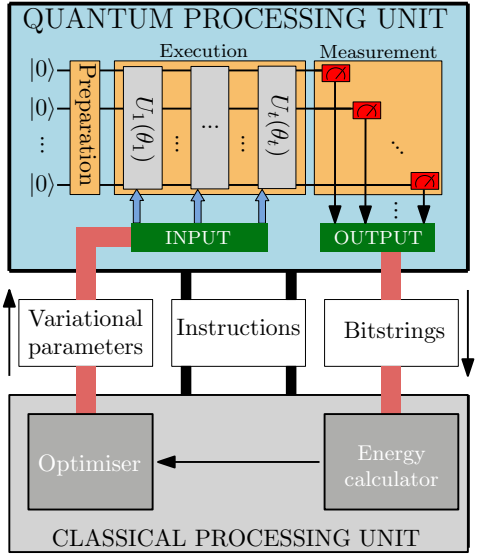}
    \caption{\textbf{Schematic diagram of hybrid quantum classical variational algorithms.} The tasks are divided between a QPU and a CPU. The CPU suggests certain parameters of a parametrised quantum circuit. This quantum circuit, along with other instructions, is then sent to and executed by the QPU. Following this, the measurement results from the QPU are sent back to the CPU, which in turn evaluates the energy to be optimised. From this result, the CPU obtains new parameters for the quantum circuit. This process continues until some convergence criterion is met.}
    \label{willsch_fig_labelv}
\end{figure}

In this article, we consider simulations of hybrid quantum classical variational algorithms. Such algorithms obtain their name from the combined usage of gate-based quantum and classical computers in a single application. The working of such algorithms is visualised in Fig.~\ref{willsch_fig_labelv}. The optimiser in the classical central processing unit (CPU) sends the parameters placed in a circuit to be executed by the quantum processing unit (QPU). The QPU prepares a problem specific initial state, executes the circuit, and sends the results to the CPU after measurement. The outcomes of the measurements, called bitstrings, are used to calculate the energy which is given to the optimiser. The optimiser then decides what new parameters will lower the energy, and the cycle continues until convergence. The CPU also controls the QPU through other instructions, e.g. the number of measurements, microwave pulses, or the time between measurements.

This article is structured as follows. 
Section \ref{willsch_sec_juqcs} contains an overview of the J\"ulich Universal Quantum Computer Simulator, which is used for the simulation of the following two hybrid quantum classical applications.
In Sec.~\ref{willsch_sec_qaoa}, we introduce the QAOA, discuss its relation to AQA, and compare both algorithms when applied to the tail assignment problem. 
In Sec.~\ref{willsch_sec_vqe}, we discuss the VQE and its application to the Heisenberg model.
Section \ref{willsch_sec_conclusion} contains our conclusions.

\section{J\"ulich Universal Quantum Computer Simulator}
\label{willsch_sec_juqcs}

In this section, we briefly describe the J\"ulich Universal Quantum Computer Simulator (JUQCS), and outline how a program for a gate-based quantum computer can be simulated using supercomputers. More details and in-depth descriptions of the implementation are given in Refs.~\citen{DeRaedt2007MassivelyParallel,DeRaedt2018MassivelyParallel,Willsch2021JUQCSGQAOA}.

The basic unit of computation for quantum computers is the qubit. Mathematically, a qubit is described by a unit vector of two complex numbers $\ket\psi = (\psi_0, \psi_1)$ fulfilling $\braket{\psi|\psi}=|\psi_0|^2+|\psi_1|^2=1$. Usually, the basis states (i.e., orthonormal basis vectors) are denoted by $\ket{0}$ and $\ket{1}$. An $N$-qubit system is defined by $2^N$ complex numbers 
\begin{eqnarray}
    \ket\psi&=&\psi_{0\ldots00}\ket{0\ldots00}+\psi_{0\ldots01}\ket{0\ldots01}+\ldots+\psi_{1\ldots11}\ket{1\ldots11},
\end{eqnarray}
where $\ket{0\ldots00},\ldots,\ket{1\ldots11}$ denote the computational basis states~\cite{NielsenChuang} and the complex coefficients $\psi_{0\ldots00},\ldots,\psi_{1\ldots11}$ again fulfil $\braket{\psi|\psi}=1$.
The $2^N$ complex coefficients in the state 
$\ket\psi$ can be written as a rank-$N$ tensor $\psi_{q_{N-1}\cdots q_1 q_0}$ where $q_j\in\{0,1\}$ denote the indices.

Since the number of coefficients grows exponentially in the number of qubits $N$, the memory requirement grows exponentially too, which makes large-scale simulations of universal quantum computers only possible on supercomputers with enough (distributed) random access memory.
To simulate for instance a quantum computer with $N=42$ qubits, the tensor $\psi_{q_{N-1}\cdots q_1 q_0}$ requires $16\times2^{42}\,\mathrm B = 64\,\mathrm{TiB}$ of memory (using double precision floating-point numbers) while the simulation of $N=21$ qubits only requires $16\times 2^{21}\,\mathrm{B}=32\,\mathrm{MiB}$ of memory.

\begin{figure}
  \centering
  \includegraphics[width=\textwidth]{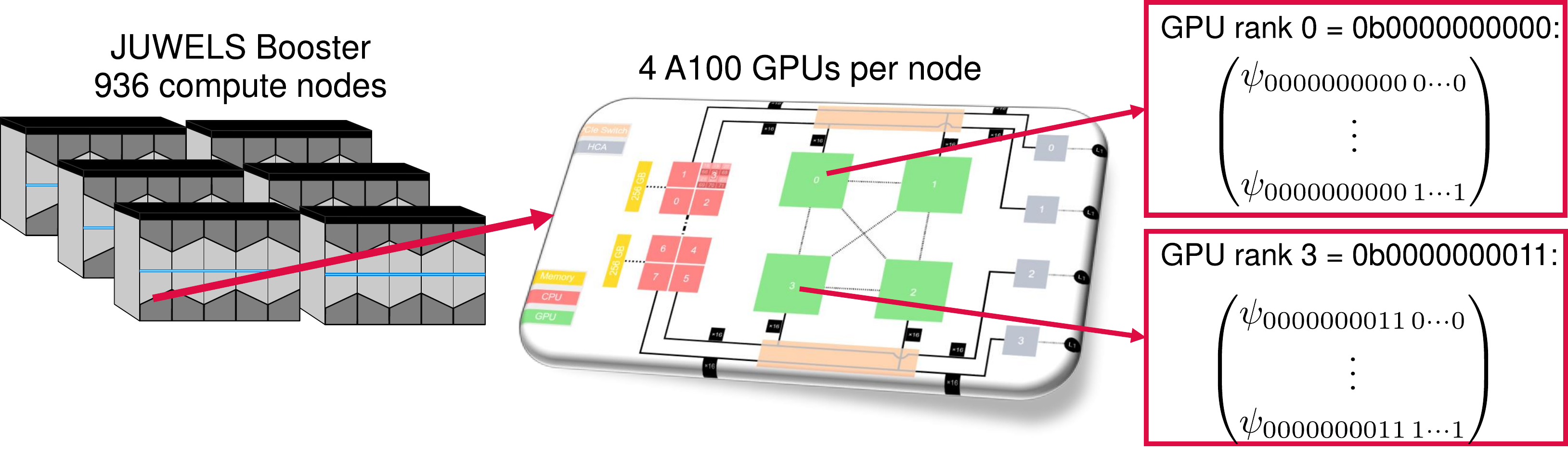}
  \caption{\textbf{Distribution of the coefficients of the state vector $\ket\psi$ across the GPUs and compute nodes of JUWELS Booster\cite{JUWELSBooster}.} Each GPU is handled by a single MPI process. For each GPU, the global qubit indices of the coefficients represent its MPI rank. For the GPUs belonging to
  MPI rank 0 and 3 this is indicated on the right (the 10 leftmost indices are the global qubit indices). The complex coefficients
  of the local qubits are stored locally on the GPUs.}
  \label{willsch_fig_mpi}
\end{figure}

The GPU version of JUQCS (JUQCS--G) distributes the complex coefficients over the memory of the participating GPUs. A sketch is shown in Fig.~\ref{willsch_fig_mpi}.
Each GPU stores a power of two coefficients, say $2^M$ coefficients, of the state vector $\ket\psi$ in its local memory. Qubits $j$ whose amplitudes $\psi_{q_{N-1}...q_{j+1}0q_{j-1}...q_0}$ and $\psi_{q_{N-1}...q_{j+1}1q_{j-1}...q_0}$ are stored on the same local memory are thus called \emph{local} qubits.
Qubits whose amplitudes are distributed over different GPUs are called \emph{global} qubits. The total number of GPUs needed is $N_{\mathrm{GPU}}=2^{N-M}$. 

Since the state vector is distributed over several GPUs, potentially also over different compute nodes, it is necessary to transfer data from one GPU to another over the network, for instance if (part of) the data is needed for a computation on another GPU or node. We use the \emph{Message Passing Interface} (MPI) to transfer the data between different GPUs.
Each MPI process controls a GPU, and its rank $r\in\{0,\ldots,N_{\mathrm{GPU}}-1\}$ is given by the global qubit indices in binary notation.

The application of a quantum gate is performed as the matrix-vector update $U\ket\psi$, where the matrix $U$ is unitary and acts only on a few-qubit subspace. For instance, the matrix corresponding to a single-qubit gate is a $2\times2$ unitary matrix $U_1 = (u^{(1)}_{qq'})$. It acts only on a single-qubit subspace and thus transforms the coefficients of $\ket\psi$ in terms of 2-component updates: Acting on qubit $j$, $U_1$ transforms the state vector $\ket\psi$ according to
\begin{align}
    \label{willsch_eq_twocomponentupdates}
    \psi_{q_{N-1}\cdots q_{j+1}qq_{j-1}\cdots q_0} \leftarrow \sum_{q'=0}^1 u^{(1)}_{qq'} \psi_{q_{N-1}\cdots q_{j+1}q'q_{j-1}\cdots q_0},
\end{align}
for $q=0,1$. Similarly, a two-qubit gate is a $4\times 4$ unitary matrix that acts on a two-qubit subspace of $\ket\psi$, and so on. 
A suitable set of one- and two-qubit gates (called a universal gate set) is sufficient for the implementation of a universal quantum computer (simulator)~\cite{Deutsch95universality, divincenzo1995twoqubitgates}.
The full gate set of JUQCS is documented in Ref.~\citen{DeRaedt2018MassivelyParallel}.

The implementation of JUQCS exploits the special structure of single-, two- and three-qubit operations. Large, dense matrices are never stored nor operated on.
It is sufficient to loop over pairs of coefficients of the state vector and to perform the 2-component updates given in Eq.~(\ref{willsch_eq_twocomponentupdates}) for the same $2\times2$ matrix $U_1$ (how $U_1$ looks like depends on the particular gate). Depending on the index of the qubit that the gate operates on, the pairs of coefficients are grouped differently. Accordingly, for two-qubit gates, the loop goes over quadruples of coefficients of $\ket{\psi}$ and the 4-component updates are performed with a $4\times4$ matrix (whose entries are determined by the particular gate).

A quantum gate acting on a global qubit requires MPI communication, because coefficients of $\ket\psi$ that need to be combined in the update are stored on different GPUs. 
Thus, to perform a single-qubit gate on a global qubit, the transfer of $2^N/2$ coefficients (i.e., half of the state vector) between pairs of GPUs is required. The communication overhead is minimised by \emph{relabelling} global and local qubits after the transfer happened once. Thus, the coefficients do not need to be transferred back again after the transformation.
Each MPI rank has to keep track of the current labelling of global and local qubits.
More details on this MPI communication scheme are given in Ref.~\citen{DeRaedt2007MassivelyParallel}.

\section{Quantum Approximate Optimisation Algorithm}
\label{willsch_sec_qaoa}

The QAOA\cite{Farhi2014QAOA} is an algorithm for gate-based quantum computers to solve combinatorial optimisation problems. Such problems can be mapped to binary optimisation problems, which are defined in terms of an $N$-bit binary string $\mathbf{x} = x_1 \dots x_N$, with the goal of finding a bitstring $\mathbf{x}$ that minimises a classical objective function $C(\mathbf{x}) : \{0, 1\}^N \rightarrow \mathbb{R}$. 

For quantum computers, a natural way of encoding a binary optimisation problem is to express its solution as the ground state (i.e., the state corresponding to the lowest energy) of the so-called Ising Hamiltonian
    \begin{equation}
        \label{willsch_eq_isinghamiltonian}
        H_\mathrm{C} = \sum_{i=0}^{N-1} h_i\sigma_i^z + \sum_{i<j} J_{ij}\sigma_i^z\sigma_j^z.
    \end{equation}
Here, $\{h_i\}$ and $\{J_{ij}\}$ denote real numbers that represent the optimisation problem instance, and $\sigma_i^z=I\otimes I\cdots I\otimes\mathrm{diag}(1,-1)\otimes I\otimes\cdots\otimes I$ (with $I$ being the $2\times2$ identity matrix) is the Pauli $z$ matrix acting on the $i^{\mathrm{th}}$ qubit subsystem.

\begin{figure}
    \centering
    \begin{equation*}
        \Qcircuit @C=1.2em @R=0.7em @!R {
                    & & & & \mbox{Repeat for $k=1,\ldots,p$ QAOA steps \hphantom{MM}} & & \\
                    & & & & \mbox{Weighting \hphantom{MMMMMMMM}} & \mbox{Mixing} & & \\
                    \lstick{ \ket{{q}_{0}}  } & \gate{H} & \qw & \gate{R^z(2\gamma_kh_0)} & \multigate{3}{\prod\limits_{i,j} e^{-i\gamma_k J_{ij}\sigma_i^z\sigma_j^z}} & \gate{R^x(2\beta_k)} & \qw & \meter \\
                    \lstick{ \ket{{q}_{1}}  } & \gate{H} & \qw & \gate{R^z(2\gamma_kh_1)} & \ghost{\prod\limits_{i,j} e^{-i\gamma_k J_{ij}\sigma_i^z\sigma_j^z}} & \gate{R^x(2\beta_k)} & \qw & \meter\\
                    \lstick{ \vdots\:\:\,  } & \vdots & & \vdots & & \vdots & & \vdots \\
                    \lstick{ \ket{{q}_{N-1}}  } & \gate{H} & \qw & \gate{R^z(2\gamma_kh_{N-1})} & \ghost{\prod\limits_{i,j} e^{-i\gamma_k J_{ij}\sigma_i^z\sigma_j^z}} & \gate{R^x(2\beta_k)} & \qw & \meter \gategroup{3}{4}{6}{5}{1em}{^\}} \gategroup{3}{6}{6}{6}{1em}{^\}}\gategroup{2}{4}{6}{6}{1.5em}{--}  \\
             }
    \end{equation*}
    \caption{\textbf{General QAOA circuit.} Initially, we bring the qubits into a uniform superposition over all states using the Hadamard gate $H$. Then, we apply $k=1,\ldots,p$ QAOA steps with parameters $\beta_1,\ldots,\beta_p$ and $\gamma_1,\ldots,\gamma_p$ (see Eq.~(\ref{willsch_eq_QAOA_state})). Each QAOA step $k$ consists of a ``weighting step'' using single-qubit $z$ rotation gates  $R^z(\varphi)=\exp(-i\varphi\sigma^z/2)$ and the set of two-qubit gates indicated in the circuit, 
    followed by a ``mixing step'' using single-qubit $x$ rotation gates 
    $R^x(\varphi)=\exp(-i\varphi\sigma^x/2)$. 
    The result after the measurement gates is used to update the variational parameters until the energy is sufficiently low (cf.~Fig.~\ref{willsch_fig_labelv}).}
    \label{willsch_fig_qaoa_circuit}
\end{figure}
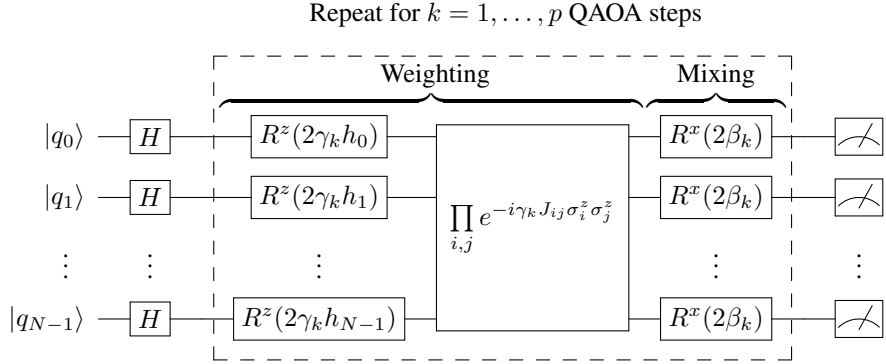

The idea of the QAOA is to solve the optimisation problem by bringing the quantum computer into a variational state 
    \begin{align}
        \label{willsch_eq_QAOA_state}
        \ket{\beta,\gamma} = \left(\prod_{k=p}^1 \exp \left[ -i \beta_k H_\mathrm{D} \right] \exp \left[ -i \gamma_k H_\mathrm{C} \right]\right) \ket{+}^{\otimes N},
    \end{align}
where $\gamma=(\gamma_1,...,\gamma_p)$ and $\beta=(\beta_1,...,\beta_p)$ are the $2p$ variational parameters to be optimised, $\ket+=(\ket0+\ket1)/\sqrt2$ is the uniform superposition of the qubit state $\ket0$ and $\ket1$, $H_C$ is the cost Hamiltonian encoding the optimisation problem (see Eq.~(\ref{willsch_eq_isinghamiltonian})), and $H_\mathrm{D} = -\sum_{i=0}^{N-1} \sigma_i^x$ is a so-called driving Hamiltonian (i.e., a sum of Pauli $x$ matrices). By finding optimal angles $\beta$ and $\gamma$ that minimise the energy expectation value $E_p(\beta,\gamma) = \bra{\beta,\gamma} H_\mathrm{C} \ket{\beta,\gamma}$, one hopes to find the ground state of $H_\mathrm{C}$. As $\ket{\beta,\gamma}$ is a linear combination of $2^N$ potential states, the idea is that optimising the $2p$ variational parameters may be much simpler than solving the original optimisation problem. However, note that in general, determining a set of optimal $\beta$ and $\gamma$ may require numerous queries to a quantum computer.
A quantum circuit to bring the quantum computer into the variational state $\ket{\beta,\gamma}$ given by Eq.~(\ref{willsch_eq_QAOA_state}) is shown in Fig.~\ref{willsch_fig_qaoa_circuit}. 

\subsection{Emulation of the QAOA}

To emulate the QAOA, JUQCS produces the variational state $\ket{{\beta,\gamma}}$ (see Eq.~(\ref{willsch_eq_QAOA_state})) by simulating the quantum circuit in Fig.~\ref{willsch_fig_qaoa_circuit} for given variational parameters $\beta$ and $\gamma$. 
For the optimisation phase of the QAOA, the energy expectation value in the variational state $E_p(\beta,\gamma)=\bra{{\beta,\gamma}}H_C\ket{{\beta,\gamma}}$ is computed by JUQCS and passed to the optimiser (cf.~Fig.~\ref{willsch_fig_labelv}; we use optimisers from the \texttt{scipy} library \cite{scipy}).
The optimiser then returns new values for the variational parameters, which are in turn used for the next simulation of the QAOA circuit, and so on.
In case the optimisation does not converge or if it converges too slowly, an additional stopping criterion of a maximum of 200 calls to JUQCS is used.

Note that in practice, the only quantity that can be estimated using a real quantum computer is the energy expectation value $E_p(\beta,\gamma)$, and not the success probability $P_{\mathrm{success}}$ (i.e., the probability of the ground state in the variational state $\ket{{\beta,\gamma}}$). 
Although we could in principle also optimise for the success probability, we consider the realistic situation and optimise for the energy only. However, the success probability is also evaluated by JUQCS to assess the quality of the solution (cf.~Fig.~\ref{willsch_fig_aqa_qaoa} below).

\subsection{Approximate Quantum Annealing}
\label{willsch_sec_AQA}
In this section, we first review how to simulate the time evolution of a quantum annealing process, and then we explain under which conditions we refer to ``approximate'' quantum annealing (AQA). Additionally, we point out the structural relation to the QAOA introduced above. A more detailed discussion of AQA and its comparison to the QAOA is given in Ref.~\citen{Willsch2021JUQCSGQAOA}.

In order to simulate a quantum annealing process, the time-dependent Schr\"odinger equation (TDSE) with $\hbar=1$, $i\partial_t\ket{\psi(t)}=H(t)\ket{\psi(t)}$, for the quantum annealing Hamiltonian $H(t)=A(t)H_\mathrm{D}+B(t)H_\mathrm{C}$ has to be solved at times $0\le t\le t_{\mathrm{anneal}}$, where $t_{\mathrm{anneal}}$ is the annealing time. Here, $H_D$ ($H_C$) is the driving (cost) Hamiltonian used in Eq.~(\ref{willsch_eq_QAOA_state}), and the so-called annealing functions $A$ and $B$ satisfy $A(0) \gg B(0)$ and $A(t_{\mathrm{anneal}}) \ll B(t_{\mathrm{anneal}})$. Therefore, we also call $H(0)\approx H_D$ and $H(t_{\mathrm{anneal}})\approx H_C$ the initial and final Hamiltonian, respectively.

The TDSE can be solved by time-stepping using the second-order Suzuki-Trotter product formula~\cite{Suzuki1993GeneralDecompositionTheoryOrderedExponentials,deraedt2004computational}
\begin{align}
    |\Psi((l+1)\tau)\rangle &=
    \left\{\exp\left[-\frac{i\tau}{2} A(l\tau)H_D\right]\right.
    \times \exp\!\left[ -i\tau B(l\tau)H_C \right]\nonumber\\
    &\ \ \ \ \!\left.\times\exp\left[-\frac{i\tau}{2} A(l\tau)H_D\right]\right\}|\Psi(l\tau)\rangle\;,
    \label{willsch_eq_pf_update}
\end{align}
where $\tau$ denotes the time step and $l=0,\ldots,n$ (such that $t_{\mathrm{anneal}}=(n+1)\tau)$.
The action of the three matrix exponentials in Eq.~(\ref{willsch_eq_pf_update}) on any state vector can be computed analytically.
The initial state is taken to be the ground state of $H_\mathrm{D}$, namely $|\Psi(0)\rangle = |+\rangle^{\otimes N}$.
Note that the structure of Eq.~(\ref{willsch_eq_pf_update}) and of the QAOA (cf.~Eq.~(\ref{willsch_eq_QAOA_state})) are essentially the same.

The idea of AQA is now to use a time step $\tau$ that is too large and a number of time steps $n$ that is too small to yield an accurate discretisation of the time evolution. Thus, we do not expect that the sequence of states $|\Psi((l+1)\tau)\rangle$ resembles the solution of the time evolution of a quantum annealing process. However, we hope that we can still generate a final state which has a relatively large overlap with the ground state of the final problem Hamiltonian $H_\mathrm{C}$ by using only a small number of matrix-vector operations. This heuristic method is motivated by empirical findings~\cite{Crooks2018PerformanceQAOAMaxCut,Brandao2018QAOAFixedControlParameters,Willsch2019BenchmarkingQAOA,Zhou2018QAOATensorFlow,Vikstal2019QAOATailAssignment} that the values for $\beta$ and $\gamma$ obtained from the optimisation were often found to approximately follow curves that are suitable as coarsely discretised quantum annealing functions $A$ and $B$.

The amounts of single-qubit and two-qubit gates required to implement either one time step of AQA, or one layer of the QAOA, are the same. Thus, for fixed $n$, (i.e., $n+1$ AQA time steps) the computational effort is equivalent to a single QAOA circuit evaluation (i.e., no optimisation) where $p=n+1$.

\subsection{Tail Assignment Problem}

The tail assignment problem \cite{Groenkvist2005TailAssignmentProblem} is a formulation of aircraft assignment problems that respects typical operational constraints such as connection times, preassigned activities, maintenance and airport curfews. 
We consider the constraint-only version of the tail assignment problem used in Ref.~\citen{Willsch2021BenchmarkAdvantage}  (other models for the tail assignment problem can be found in Refs.~\citen{Vikstal2019QAOATailAssignment,Martins2020TailAssignmentProblemQA,svensson2021LargeILPBranchPrice}). In this form, the problem can be formulated as an exact cover problem, which is an NP-complete problem\cite{Karp1972KarpsNPCompleteProblems} from the class of set covering and partitioning problems. The corresponding classical objective function to be minimised reads
\begin{align}
    C(\mathbf{x}) = \sum_{f=1}^{F}\left(\sum_{i=1}^N \mathbf{A}_{if}x_i - 1\right)^2.
\end{align}
Here, $\mathbf{A}\in\{0,1\}^{N\times F}$ is a Boolean matrix that defines the exact cover problem instance. In the context of aircraft assignment, each of the $F$ columns of $\mathbf{A}$ represents one flight, and each of the $N$ rows represents a given route. Thus, $\mathbf{A}_{if}=1$ if and only if flight $f$ is contained in route $i$. The solution $\mathbf{x} = x_1 \dots x_N$ therefore selects a set of aircraft routes from $\mathbf{A}$ such that each flight is covered exactly once (see Ref.~\citen{Willsch2021BenchmarkAdvantage}, which also contains the explicit expression for $H_\mathrm{C}$ in Eq.~(\ref{willsch_eq_isinghamiltonian}) as a function of $\mathbf A$).
We consider the same $N=30,\ldots,40$ qubit problems with $F=472$ flights that have been used in Ref.~\citen{Willsch2021JUQCSGQAOA}.

\subsection{Results}

\begin{figure}
  \flushright
  \includegraphics[width=.85\textwidth]{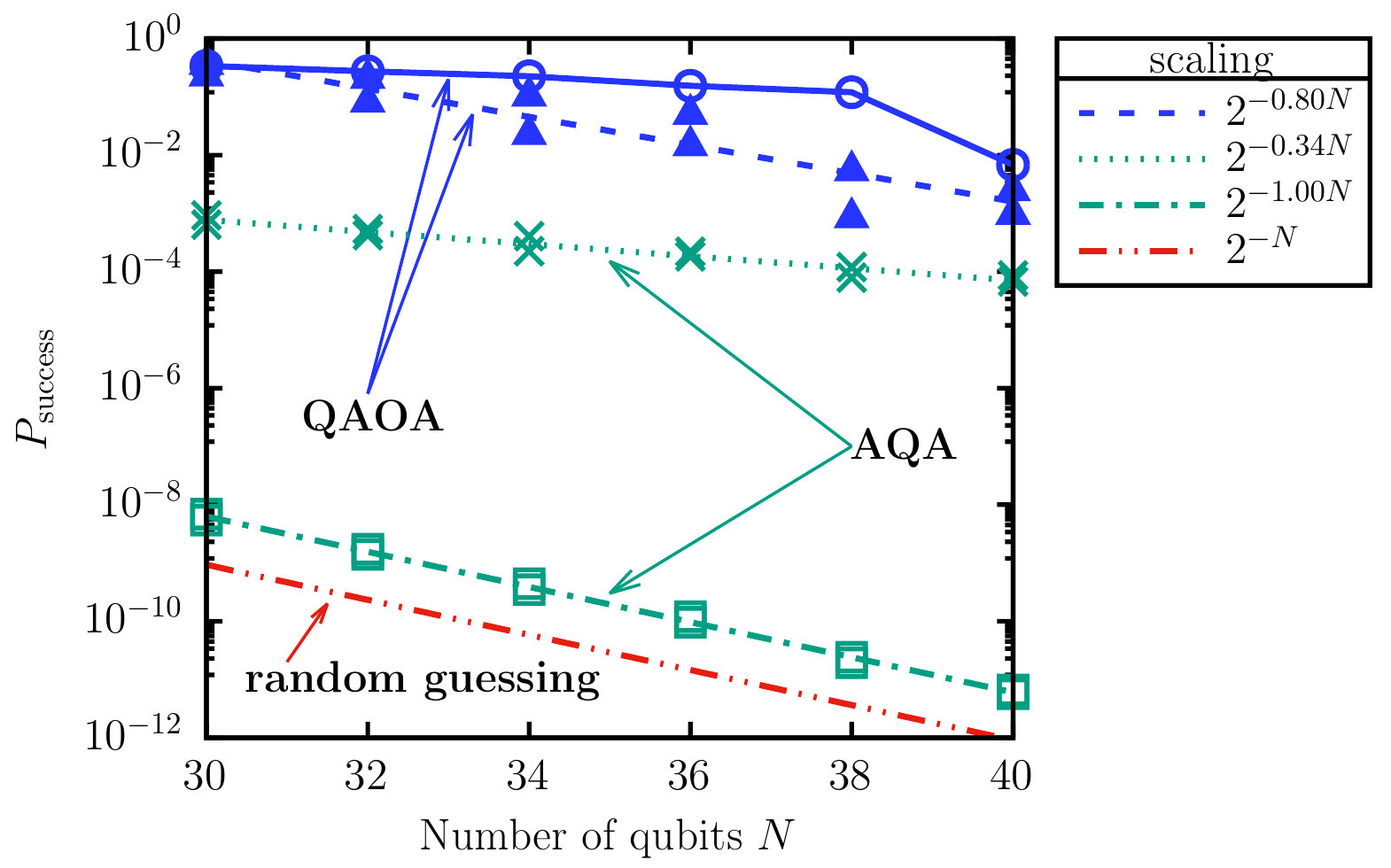}
  \caption{\textbf{Success probabilities $P_\mathrm{success}$ obtained from AQA and the QAOA, in comparison to random guessing, for different problem sizes $N$.} The success probabilities obtained with the QAOA are shown for a case in which the variational parameters are separately optimised for each problem size $N$ (blue circles, $p=7$), and for a case in which fixed variational parameters are taken from a previous optimisation of the $N=30$ case (blue triangles, $p=6$). The success probabilities obtained with AQA are shown for $\tau=0.8$ (green crosses) and $\tau=0.1$ (green squares). 
  The dashed, dotted, and dash-dotted lines represent fits to the average success probabilities (in these cases, two instances per problem size have been simulated).
  The red dash-dot-dotted line represents the probability $1/2^N$ to find the solution by random guessing.
  The solid line is a guide to the eye.}
  \label{willsch_fig_aqa_qaoa}
\end{figure}
In this section, we summarise some representative QAOA and AQA results. In Fig.~\ref{willsch_fig_aqa_qaoa}, we plot the success probability $P_{\mathrm{success}}$ as a function of the problem size $N$ for two versions of AQA and the QAOA. The AQA results are for $n=5$ steps with time step $\tau=0.1$ (green squares) and $\tau=0.8$ (green crosses). The QAOA results are for standard QAOA (i.e., the variational parameters are optimised individually for each problem instance, initialised from AQA with $n=6$ and $\tau=0.4$; blue circles), and for pre-optimised QAOA (i.e., the variational parameters are optimised for a $30$-qubit problem, initialised from AQA with $n=5$ and $\tau=0.2$, and these same values are reused for all larger problem instances; blue triangles). Further results and additional details can be found in Ref.~\citen{Willsch2021JUQCSGQAOA}.

Both QAOA and AQA obtain better success probabilities than can be found by random guessing (red dash-dot-dotted line), but for the QAOA they are overall higher than for AQA. This is reasonable since the variational QAOA parameters have been initialised with AQA results.
More important, however, is the scaling behaviour as a function of the problem size $N$.
For AQA with $\tau=0.1$ (dash-dotted line), the scaling is equal to the scaling of random guessing ($1/2^N$).
The scaling of the QAOA with pre-optimised variational parameters (dashed line) is only marginally better. However, both AQA with $\tau=0.8$ (dotted line) and standard QAOA (solid line) show considerably good scaling behaviour (albeit still exponential).

Merely the particular QAOA data point for the largest problem size (the blue circle at $N=40$) falls out of this scaling behaviour. Here, the maximum number of $200$ JUQCS-calls (and thus the limit of the computational budget allocated for this job) was reached.
This represents the fact that the QAOA optimisation procedure requires a significant amount of computational work on many nodes of the supercomputer. 

This is a noteworthy advantage of AQA, as it can extend the good scaling also to $N=40$ without requiring any involved optimisation procedure once a good value for $\tau$ is found.
Interestingly, the regime of ``good'' values appears to be for large time steps $\tau$, where AQA is actually not a genuine discretisation of quantum annealing anymore.

\section{Variational Quantum Eigensolver}
\label{willsch_sec_vqe}

An interesting and common problem is to find the ground state energy of a Hamiltonian $H$. As this is a hard problem to solve on classical computers, quantum computers are considered as an alternative route to solve the problem. When the ground state $\ket{\psi_0}$ can be prepared on a quantum computer, the expectation value
\begin{equation}
\braket{H}_0 =  \frac{\bra{\psi_0}H\ket{\psi_0}}{\braket{\psi_0|{\psi_0}}}
\end{equation}
gives the ground state energy. The task is to find this ground state energy by initialising the quantum computer in an arbitrary initial trial state $\ket{\psi_t}$. The variational theorem states that $\braket{H}_t$ for any $\ket{\psi_t}\neq \ket{\psi_0}$ is an upper bound to the ground state energy. This means that
\begin{equation}
\braket{H}_t \geq E_0, \label{willsch_vqeeq1}
\end{equation}
where the equality sign holds if and only if $\ket{\psi_t}= \ket{\psi_0}$. In Eq.~(\ref{willsch_vqeeq1}), $E_0$ is the lowest eigenvalue or the ground state energy of the physical system described by $H$.

The VQE is an implementation of the variational principle on a quantum computer. Its working is illustrated in Fig.~\ref{willsch_fig_labelv}. The VQE derives its name from the fact that it was used to solve for the lowest eigenvalue of a Hamiltonian \cite{willsch_Peruzzo2014}.

\subsection{Heisenberg model}
The Hamiltonian for the one-dimensional Heisenberg model is given by
\begin{equation}
    H = \sum_{i>j}^N \Big({J}_{ij}^{xx} \sigma_i^x \cdot \sigma_j^x+{J}_{ij}^{yy} \sigma_i^y \cdot \sigma_j^y+{J}_{ij}^{zz} \sigma_i^z \cdot \sigma_j^z \Big ),
\end{equation}
where $N$ denotes the number of spins, $\sigma^x,\sigma^y,$ and $\sigma^z $ are the Pauli matrices, and $J_{ij}$ is the coupling interaction between the $i^{\text{th}}$ and $j^{\text{th}}$ spins. We use units such that $\hbar=1$ and the $J$'s are dimensionless. If $J_{ij} = 1$ for all $i,j=1,\dots,N$, we call $H$ the \textit{isotropic antiferromagnetic} model Hamiltonian. We consider bipartite spin lattices in our simulations.

\subsubsection{The ansatz}
The variational principle allows any ansatz to be used to solve the ground state energy problem. However, not every ansatz can express the ground state of $H$. An ansatz\cite{jattana2022assessment} that can find a reasonable approximation to the ground state energy is given by 
\begin{equation}
U(\bm \theta) = \Big [\prod_{ l =N-1 }^{1} \prod_{k = N}^{ l+1} U_{lk}(\theta_{lk}) \Big ] \Big [\prod_{ l =N-1 }^{1} \prod_{k = N}^{ l+1} U_{kl}(\theta_{kl}) \Big ], \label{willsch_man1}
\end{equation}
where 
\begin{equation}
      U_{pq}(\theta_{pq}) =
    \begin{cases}
      e^{-i\theta_{pq} \sigma_p^y\sigma_q^x } & \text{if $p=N$ or $q=N$,}\\
      e^{-i\theta_{pq} \sigma_p^y\sigma_q^x\sigma_N^z } & \text{otherwise.}
    \end{cases}   
\end{equation}

If an ansatz can express the ground state, it is still challenging to optimise the parameters such that $\ket{\psi_t(\bm \Theta)} = \ket{\psi_0}$, where $\bm \Theta$ are the optimised parameters. In addition to the ansatz, it is important to specify the initial state (say $\ket{\Psi_0}$) of the quantum computer. The ansatz then acts on the initial state, i.e. $\bm U(\bm \theta) \ket{\Psi_0}$, with a set of initial parameters to give the initial energy. The task of the VQE is to lower this energy until it equals the ground state energy. A better choice of initial energy can improve the performance of the VQE. To obtain a better (than random) initial energy for the Heisenberg model, it helps to set the initial state to the N\'{e}el state and the initial parameters to zero. On a lattice with an even number of spins, the N\'{e}el state is prepared by setting adjacent spins anti-parallel. On a quantum computer, and for a one dimensional model, this is achieved by setting the qubits with the odd (even) indices to $\ket{0}$ ($\ket{1}$), or vice versa.

\subsubsection{Results}

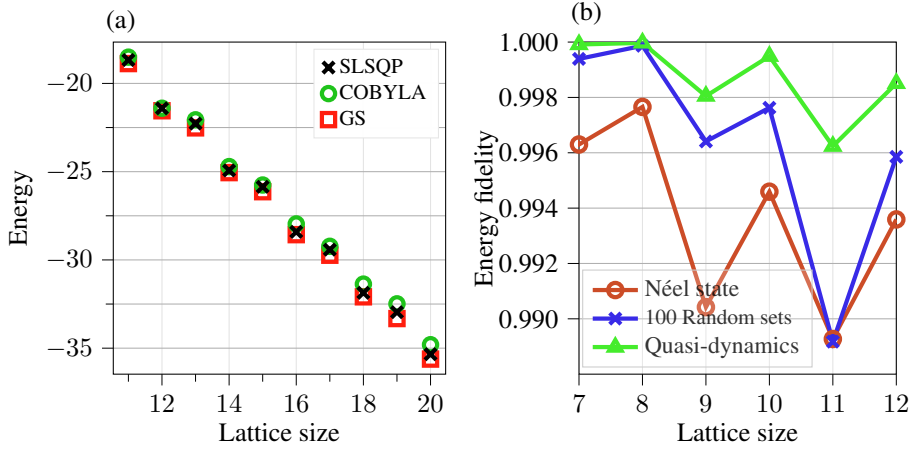
\begin{figure}
    \centering
\begin{tikzpicture}

\definecolor{color1}{rgb}{0.12156862745098,0.766666666666667,0.105882352941177}
\definecolor{color0}{rgb}{1,0.118039215686275,0.0549019607843137}
\node at (0.1,4.7) {(a)};
\begin{axis}[
legend cell align={left},
legend style={fill opacity=0.99, draw opacity=1, text opacity=1, draw=white!90!black},
reverse legend,
tick align=outside,
tick pos=left,
legend style={nodes={scale=0.8, transform shape}}, 
x grid style={white!89.0196078431373!black},
y grid style={white!89.0196078431373!black},
xmin=10.55, xmax=20.45,
xtick style={color=black},
y grid style={white!69.0196078431373!black},
ymin=-36.472771363, ymax=-17.657816017,
ytick style={color=black},
width=6cm,
ylabel = {Energy},
height=6cm,
xmajorgrids,
minor tick num = 1,
yminorgrids,
ymajorgrids,
xlabel = {Lattice size}
]
\addplot [ultra thick, mark size = 2.5pt, draw=color0, fill=color0, mark=square, only marks]
table{%
x  y
11 -18.875745452
12 -21.54956364
13 -22.51833732
14 -25.054198136
15 -26.134670288
16 -28.569185444
17 -29.73408066
18 -32.090996348
19 -33.321953816
20 -35.61754612
};
\addlegendentry{GS}
\addplot [ultra thick, draw=color1, mark size = 2.5pt, fill=color1, mark=o, only marks]
table{%
x  y
11 -18.51304126
12 -21.40671106
13 -22.06295936
14 -24.71172702
15 -25.75645439
16 -27.9616094
17 -29.22878214
18 -31.37082272
19 -32.48905695
20 -34.80389974
};
\addlegendentry{COBYLA}
\addplot [ultra thick, mark size = 3pt, draw=black, fill=black, mark=x, only marks]
table{%
x  y
11 -18.67322287
12 -21.41145504
13 -22.27494651
14 -24.929709
15 -25.85808528
16 -28.41071591
17 -29.41870365
18 -31.86377001
19 -32.96127721
20 -35.34168862
};
\addlegendentry{SLSQP}
\end{axis}

\end{tikzpicture}
\begin{tikzpicture}

\definecolor{color1}{rgb}{0.22156862745098,0.166666666666667,0.905882352941177}
\definecolor{color0}{rgb}{0.8,0.298039215686275,0.1549019607843137}
\definecolor{color2}{rgb}{0.272549019607843,0.927450980392157,0.172549019607843}
\node at (0.1,4.8) {(b)};
\begin{axis}[
ylabel = {Energy fidelity},
legend cell align={left},
legend style={fill opacity=0.2, draw opacity=1, text opacity=0.8, at={(0.03,0.03)}, anchor=south west, draw=white!80!black},
legend style={at={(0.01,0.02)},anchor=south west, nodes={scale=0.9, transform shape}},
tick align=outside,
tick pos=left,
ymax = 1.0, ymin = 0.988,
x grid style={white!89.0196078431373!black},
xmin=7, xmax=12,
xtick = {4,5,6,7,8,9,10,11,12},
xtick style={color=black},
y grid style={white!69.0196078431373!black},
ytick style={color=black},
width=5.8cm,
height=6cm,
yminorgrids,
ymajorgrids,
xmajorgrids,
ytick = {0.990, 0.992, 0.994,0.996, 0.998, 1.000},
yticklabels ={0.990, 0.992,0.994, 0.996,0.998, 1.000},
xlabel = {Lattice size}
]
\addplot [ultra thick, mark=o, mark size = 2.7pt, draw=color0, colormap/viridis]
table{%
x                      y
6 0.999995997875644
7 0.996299283346753
8 0.997656733775885
9 0.99042189713744
10 0.994593588514171
11 0.989270750523999
12 0.993591118488189
};
\addlegendentry{N\'{e}el state}
\addplot [ultra thick, mark=x, mark size = 3pt, draw=color1, colormap/viridis]
table{%
x                      y
6 0.9999999
7 0.999392303106875
8 0.999867098774961
9 0.996410171462558
10 0.997625644805607
11 0.989174981016974
12 0.995852211150952
};
\addlegendentry{\small{100 Random sets}}
\addplot [ultra thick, mark=triangle, mark size = 2.7pt, draw=color2, colormap/viridis]
table{%
	x                      y
	7 0.999912842322556
	8 0.99997149765541
	9 0.998049122087149
	10 0.999488979571436
	11 0.996238174254645
	12 0.998508827203549
};
\addlegendentry{Quasi-dynamics}
\end{axis}
\end{tikzpicture}
    \caption{(a) Optimised variational energies obtained using gradient based (SLSQP) and gradient free (COBYLA) algorithms in comparison to the ground state energies for different lattice sizes. (b) Comparison between the energy fidelities obtained using the N\'{e}el initial state, best of $100$ random sets of parameters, and quasi-dynamics.}
    \label{willsch_fig_label1}
\end{figure}

We find the ground state energy of isotropic Heisenberg rings of length $11$ to $20$ using two different optimisers. We used the gradient based (free) optimiser SLSQP (COBYLA) \cite{Powell1994, scipy, Kraft}. The results are compared against the ground state energy found by the Lanczos algorithm and are shown in Fig.~\ref{willsch_fig_label1}(a). We observe, as expected, that using different optimisers leads to different results. SLSQP performs better in all cases but the difference in obtained energies appears to grow with increasing lattice sizes.

\subsection{VQE and quasi-dynamics}
When using the VQE, the variational quantum state as a function of $m$ parameters is written as  
\begin{equation}
\ket{\psi} = U(\bm \theta)  \ket{\Psi_0}= U_m(\theta_m)\ldots U_1(\theta_1) \ket{\Psi_0}. \label{eq1a}
\end{equation}
If $\mathbb{U}(\bm \Theta)$ ($U(\bm \theta)$) represents the unitary operators corresponding to the optimised (unoptimised) numeric values of the parameters, then the VQE can be said to perform the task $U(\bm \theta) \to \mathbb{U}(\bm \Theta)$. After the optimiser signals convergence, the state obtained is
\begin{equation}
\ket{\Psi_1} = \mathbb{U}(\bm \Theta) \ket{\Psi_0}. \label{eq1p}
\end{equation}
The idea is to use this state as an initial state for another round of VQE optimisation. To do so, substitute Eq.~(\ref{eq1p}) in Eq.~(\ref{eq1a}) repeatedly ($p$ times), such that the final state is
\begin{equation}
\ket{\Psi_p} =  \mathbb{U}_p(\bm \Theta_p) \ldots \mathbb{U}_1(\bm \Theta_1) \ket{\Psi_0}. \label{willsch_eqse}
\end{equation}
The procedure represented by Eq.~(\ref{willsch_eqse}) is termed \emph{quasi-dynamical evolution} \cite{man2}. 
A suitable choice of initial parameters for each $U(\bm \theta)$ is  $\bm \theta = [ 0,\dots,0 ]$. This step is necessary in order to preserve the progress made until then. By avoiding random initialisation we also avoid potential barren plateaus \cite{McClean2018}. Similar to the QAOA, the performance of the quasi-dynamics is expected to improve as $p$ is increased.

\subsubsection{Results}
We perform the quasi-dynamical evolution on isotropic rings of lengths $7$ to $12$. We set the threshold for the quasi-dynamics such that the process stops when the energy at step $p$ can no longer be lowered below $10^{-4}$ compared to the $(p-1)^{\text{th}}$ value. The results are compared against the final energy obtained when (1) the N\'{e}el initial state and (2) one hundred random sets of initial parameters were used. We define the energy fidelity as the ratio of the variational energy obtained upon convergence of the optimiser to the ground state energy. For (2), we plot only the highest energy fidelity obtained from all cases. The results are shown in Fig.~\ref{willsch_fig_label1}(b). We observe that the quasi-dynamical evolution can find a better energy fidelity.

In our results we used the same $\bm U(\bm \theta)$ (from Eq.~(\ref{willsch_man1})) for each of the $p$ iterations in Eq.~(\ref{willsch_eqse}). While it improves the results for the cases we tested, such an elementary choice may not always converge to the ground state energy. Equation~(\ref{willsch_eqse}) allows the use of any arbitrary $\bm U(\bm \theta)$; however, it remains an open question which choices of $\bm U(\bm \theta)$ will find the ground state energy for problems in general. 

We observed that the number of random restart cases that found an energy better than the energy found when starting from the N\'{e}el state dropped as the lattice size is increased (data not shown). Since the ansatz has more parameters for larger lattices, one solution is to increase the number of restarts. However, this approach can be very resource expensive. Thus, random initialisation is not a suitable strategy. It is crucial to have a good combination of ansatz and initial parameters to have a reasonable chance of finding the ground state energy.

\section{Conclusions}
\label{willsch_sec_conclusion}

In this article, we have discussed two popular hybrid quantum classical algorithms, namely the QAOA and the VQE. Both have been simulated on JUWELS Booster using JUQCS--G. 

For the QAOA, we found that the best scaling of the success probability can be obtained when the full variational optimisation procedure is carried out. However, due to tremendous computational resources required for simulations with $N=40$ qubits, it was not possible to reach convergence for the largest problem size. Therefore, we considered AQA as an alternative algorithm with only a single parameter $\tau$ to solve this problem. Although slightly worse in terms of actual success probabilities, AQA could extend the same promising scaling behaviour also to the largest problems. Interestingly, the resulting optimal $\tau$ was so large that it put AQA into a regime where it does not accurately describe a discretised quantum annealing process anymore.

For the VQE, we found that a suitable choice of an ansatz coupled with a good choice of initial parameters is critical for finding the ground state energies. For the Heisenberg model, one such choice is the N\'{e}el state with all variational parameters initialized to zero. However, which choices are helpful in general is an open question. We introduced and tested quasi-dynamical evolution, a technique which builds upon the VQE and improves its performance. An open question for the quasi-dynamics is the choice of $\bm U$ in each substitution which guarantees finding the ground state energy.

\section*{Acknowledgements}

The authors gratefully acknowledge the Gauss Centre for Supercomputing e.V.
(www.gauss-centre.eu) for funding this project by providing computing time
on the GCS Supercomputer JUWELS at J\"ulich Supercomputing Centre (JSC).
We would like to thank M. Svensson for providing the exact cover problem instances.
D.W. and M.W. acknowledge support from the project J\"ulich UNified Infrastructure for Quantum computing (JUNIQ) that has received funding from the German Federal Ministry of Education and Research (BMBF) and the Ministry of Culture and Science of the State of North Rhine-Westphalia. M.J.  acknowledges support from the project OpenSuperQ (820363) of the EU Quantum Flagship.

\bibliographystyle{nic}
\bibliography{database}

\end{document}